\documentclass[preprint]{revtex4}
\usepackage{epsfig,latexsym,amssymb}
\usepackage{latexsym}
\usepackage{amssymb}
\usepackage{amsfonts}
\usepackage{comment}
\usepackage{graphicx}
\usepackage{verbatim}

\newcommand{\be}{\begin{equation}}
\newcommand{\ee}{\end{equation}}
\newcommand{\bea}{\begin{eqnarray}}
\newcommand{\eea}{\end{eqnarray}}

\begin{document}

\title{Axions and Dark Matter}

\author{Qiaoli Yang$^{1,2,}$\footnote{qiaoli\_yang@hotmail.com}}
\affiliation{$^{1}$Department of Physics, Jinan University, Guangzhou 510632, China\\$^{2}$School of Physics, Huazhong University of Science and Technology, Wuhan 430074, China}

\begin{abstract}
Dark matter constitutes about $23\%$ of the total energy density of the universe but its properties are still little known besides that it should be composed by cold and weakly interacting particles. Many beyond standard model theories can provide proper candidates to serve as dark matter and the axion introduced to solve the strong CP problem turns out to be an attractive one. In this paper, we briefly review several important features of the axion and the axion dark matter.
\end{abstract}

\date\today
\maketitle

\section{Introduction}
Cosmological and astronomical observations indicate that the majority of matter in the universe is composed by some beyond standard model particles \cite{2014A&A...571A..16P}. These dark matter particles are believed to be weakly interacting, stable, and cold. The axion is originally introduced to explain the strong CP problem, but later it turns out that the cosmic axions created from the re-alignment mechanism and/or the decay of cosmic topological defects can be a substantial fraction of dark matter.

\section{The Strong CP Problem}
The standard model of particle physics successfully explains a variety of experimental results around globe up to TeV energy scale. The theory describes the elementary particles using quantized relativistic fields and the Lagrangian density of the respective fields can be cataloged into four parts: the Yang-Mills part ${\cal L}_Y$, the Weyl-Dirac part ${\cal L}_W$, the Higgs part ${\cal L}_H$, and the Yukawa coupling part ${\cal L}_{Yu}$.

The Yang-Mills part ${\cal L}_Y$ which describes elementary spin-1 bosons mediating the fundamental interactions is written as:
\begin{equation}
{\cal L}_Y=-{1\over4g_3^2}\sum G^A_{\mu\nu}G^{\mu\nu A}-{1\over4g_2^2}\sum F^a_{\mu\nu}F^{\mu\nu a}-{1\over4g_1^2}B_{\mu\nu}B^{\mu\nu},
\end{equation}
where $g_i$ are dimension-less constants that determine the respective interaction strength, $A=1...8$, and $a=1...3$. $G_{\mu\nu}^A$, $F_{\mu\nu}^a$ and $B_{\mu\nu}$ are the field strength tensors corresponding to the SU(3), the SU(2), and the U(1) group respectively.

The Weyl-Dirac part ${\cal L}_W$ describes the spin-1/2 fundamental fermions (the quarks and the leptons) and their couplings to the gauge particles. Since the left-handed fermions are SU(2) doublets and the right-handed fermions are SU(2) singlets, the standard model is chiral. The quarks are SU(3) triplets and the leptons are SU(3) singlets so the leptons do not participate the strong interactions. One can write the Weyl-Dirac Lagrangian as:
\begin{equation}
{\cal L}_W=L_i^{\dagger} \sigma^{\mu}D_{\mu}L_i+{\bar e}^{\dagger}_i\sigma^{\mu}D_{\mu}{\bar e}_i+Q^{\dagger}_i\sigma^{\mu}D_{\mu}Q_i+\bar u^{\dagger}_i\sigma^{\mu}D_{\mu}\bar u_i+\bar d~^{\dagger}_i\sigma^{\mu}D_{\mu}\bar d_i~~,
\end{equation}
where $D_{\mu}$ are the covariant derivatives of the respective fermion fields. For example, $D_{\mu}\bar e_i=(\partial_{\mu}+iB_{\mu})\bar e_i$, et al.

The Higgs part ${\cal L}_H$ is composed by the Higgs doublet: $H={H_1\choose H_2}$. The Lagrangian is:
\begin{equation}
{\cal L}_H=(D_{\mu}H)^{\dagger}(D^{\mu}H)+{\mu}^2H^{\dagger}H-\lambda(H^{\dagger}H)^2~~.
\end{equation}

Finally the Yukawa part ${\cal L}_{Yu}$ can be written as:
\begin{equation}
{\cal L}_{Yu}=iL_i^T\sigma_2\bar e_jH^*Y_{ij}^e+iQ_i^T\sigma_2\bar d_jH^*Y_{ij}^d+iQ^T_i\sigma_2\bar u_j\tau_2 HY^u_{ij}+c.c.\label{yukawa}
\end{equation}
where $Y_{ij}^k$ are $3\times 3$ matrices of the respective Yukawa couplings. The matrix of the leptons can be always written as a real diagonal matrix in an expense of mixing the lepton fields. For the quark part, there are two matrices appearing in the Lagrangian thus one finds:
\begin{equation}
{\cal L}_{Yu~Quark}=iQ_i^T\sigma_2 y_{ii}^d \bar d_iH^*+iQ_i^T\sigma_2U_{ji}y_{jj}^u\bar u_j\tau_2 H+c.c.
\end{equation}
where $y_{ii}^{d,u}$ are real numbers, and $U_{ji}$ is the Cabibbo-Kobayashi-Maskawa matrix which has three mixing angles plus one phase.

The three gauge couplings $g_1, g_2$, and $g_3$ combining with the four parameters of the CKM matrix, the nine masses of the fermions, the $\mu$ term and the $\lambda$ term in the Higgs part and one additional parameter, the QCD vacuum angle, form the complete parameter space of the standard model of particle physics.

Let us now investigate the QCD vacuum angle. As the vacuum is the lowest energy state, one of the field configurations corresponding to the vacuum is $A_{\mu}=0$. Therefore, after a gauge transformation, $A_{\mu}=(i/g) U\partial_{\mu}U^{\dagger}$ are the general field configurations of vacuum. In the following discussions let us use a gauge $A_0=0$ for convenience, thus the matrix $U$ can be written as a time independent matrix: $U=U(\vec x)$. As not every $U$ can be smoothly deformed into each other without passing through non-zero energy barriers, the $U$s can be catalogued by their topological property of mapping or their winding numbers. The winding number is defined as:
\begin{equation}
n=-{1\over 24\pi^2}\int d^3 x\epsilon ^{ijk}Tr[(U\partial_iU^{\dagger})(U\partial_jU^{\dagger})(U\partial_kU^{\dagger})].
\end{equation}

Vacuum configurations with different winding numbers are separated by energy barriers, but they can tunnel into each other due to the instantons. The tunneling amplitude between two vacuum states is: $<n2|H|n1>\sim e^{-S}$, where $S$ is the Euclidean action with the two vacuum as its boundary and $ni$ denotes their winding numbers respectively. The two sates remain degenerate if the action $S$ is infinite, however field configurations giving rise to a finite action $S$ exist. The configurations with boundary $|n1>$ at $t=-\infty$, $|n2>$ at $t=+\infty$ and $n2-n1=1$ are called the instantons. Due to the tunneling effect, the physical QCD vacuum is a superposition of the vacuum configurations with all different winding numbers. Let us denote the physical vacuum as $|\omega>$, and consider a gauge transformation $T$: $T|n>=|n+1>$. Applying the transformation $T$ on the physical vacuum, one finds $T|\omega>=e^{i\theta}|\omega>$ which means that the transformation changes the physical vacuum at most by a pure phase factor. As the physical vacuum includes all possible winding number states, we have $|\omega>=\sum_n e^{in\theta}|n>$. Therefore the argument gives rise to a new parameter $\theta\in [0,2\pi]$, which is the $\theta$ parameter of the QCD vacuum.

The $\theta$ vacuum has observable effects. Considering the vacuum to vacuum transition amplitude:
\begin{eqnarray}
<\theta_1|e^{-Ht}|\theta>&=&\sum_{n_1}\sum_ne^{-i(n_1\theta_1-n\theta)}<n_1|e^{-Ht}|n>\nonumber\\&=&\sum_{n_1}e^{-in_1(\theta_1-\theta)}\sum_{n_1-n}\int [DA_{\mu}]_{n_1-n}exp[-\int d^4xL-i(n_1-n)\theta]\nonumber\\&=&\delta(\theta_1-\theta)\int [DA_{\mu}]_qexp[-\int d^4x(L+{\theta\over 32\pi^2}F^a_{\mu\nu}\tilde F^{\mu\nu a})] ,
\end{eqnarray}
where $\int[DA_{\mu}]_{n_1-n}$ is the functional integration with respect to gauge configurations $n_1-n=q$. An effective interaction term is presented: $\theta/32\pi^2F^a_{\mu\nu}\tilde F^{a\mu\nu}$ due to the $\theta$ vacuum. Since $F\tilde F$ is CP odd, QCD will be not CP invariant when $\theta\ne 0$.

The following transformation:
\begin{eqnarray}
\theta F\tilde F\to (\theta-\sum \alpha_i)F\tilde F\\
\bar q_im_iq_i\to \bar q_ie^{(i\alpha_i \gamma_5/2)}m_ie^{(i\alpha_i\gamma_5/2)}q_i
\end{eqnarray}
can annihilate the $\theta$ term if $\sum \alpha_i=\theta$. However, the expense is the additional mass terms:
\begin{equation}
\sum (m_icos \alpha_i \bar q_i q_i+m_i sin \alpha_i \bar q_i\gamma_5 q_i).
\end{equation}
Thus the parameter determines the CP violation of the QCD is $\bar \theta=\theta+arg detM$. The $\bar\theta$ term gives rise to a contribution to the neutron electric dipole moment. Baluni and Crewther et al. calculate that the dipole momentum is $d\sim 10^{-16}\bar \theta{\rm e cm}$ but the upper boundary of the neutron electric dipole moment is $10^{-24}{\rm e cm}$. Therefore $|\bar\theta|<1.2*10^{-9}$ which is extremely small.  Why $\bar \theta$ being such a small number is the strong CP problem.

To solve the strong CP problem, there are several popular possibilities:

1. The ultraviolet up quark mass is zero. However, the lattice QCD simulations suggest a non-zero u quark mass.

2. In our universe the $\bar \theta$ term happens to be very small.

3. The Peccei-Quinn mechanism \cite{Peccei:1977hh}. If $\bar \theta$ has a dynamical counterpart, it combined with the counterpart will naturally relax to a value that minimizes the total energy, which is zero.

A proof that the energy is minimized when $\bar \theta=0$ is given by C. Vafa and E. Witten \cite{Phys. Rev. Lett. 53. 535(1984)}. Considering the path integral of the QCD action in the Euclidean space, the QCD Lagrangian is: ${\cal L}=-1/4g^2Tr(G_{\mu\nu}G_{\mu\nu})+\sum \bar q_(D_{\mu}\gamma_{\mu}+m_i)q_i+i\bar\theta/32\pi^2Tr(G_{\mu\nu}\tilde G_{\mu\nu})$. After integration of the fermions one finds:
\begin{eqnarray}
e^{-VE}=\int DA_{\mu}det(D_{\mu}\gamma_{\mu}+M)e^{\int d^4x[1/4g^2 Tr G_{\mu\nu}G_{\mu\nu}-i\bar\theta/32\pi^2G_{\mu\nu}\tilde G_{\mu\nu}]}~~.
\end{eqnarray}
In QCD $det(D_{\mu}\gamma_{\mu}+M)$ is positive as $(\gamma_{\mu}i D_{\mu})$ is hermitian in the Euclidean space and $\gamma_5$ anti-commutes with $\gamma_{\mu}$. So for every (imaginary) eigenvalue of $\gamma_{\mu} D_{\mu}$, there is an eigenvalue with an opposite sign. Note that $iG\tilde G$ is pure imaginary, so it only reduces the total value of the path integral. Therefore if $\bar\theta$ is not zero, the system energy is higher.

If the solution 3 applies, the Lagrangian can be written as:
\begin{eqnarray}
{\cal L}&=& -1/4g^2Tr(G_{\mu\nu}G_{\mu\nu})+\sum \bar q_(D_{\mu}\gamma_{\mu}+m_i)q_i\nonumber\\&+&\bar\theta/32\pi^2TrG_{\mu\nu}\tilde G_{\mu\nu}+1/2\partial_{\mu}a\partial^{\mu}a+a/(f_a32\pi^2)TrG_{\mu\nu}\tilde G_{\mu\nu}+...~~,\label{eq232}
\end{eqnarray}
where the field $a$, the axion, is the counterpart of the $\bar\theta$ term and consequently the strong CP problem is solved.

\section{Axion Models}
There are several mile-stone models of the axion. The earliest one is the Peccei-Quinn-Weinberg-Wilczek (PQWW) axion \cite{Weinberg:1977ma, Wilczek:1977pj} in which model an additional Higgs doublet is added with a $U_{PQ}(1)$ symmetry breaking scale of order $250$GeV, the electro-weak scale. Experiments have ruled out this model. J. Kim et. al. separate the electro-weak scale and the $U_{PQ}(1)$ symmetry breaking scale and assume that the latter is much higher than $250$GeV. The models with a symmetry breaking scale much higher than the electro-weak scale are known as the "invisible axions". The two major types of the invisible axions are the Kim-Shifman-Vainshtein-Zakharov (KSVZ) axion and the Dine-Fishler-Srednicki-Zhitnitskii (DFSZ) axion.

The KSVZ \cite{Kim:1979if,shifman:1979ifd,Zhitnitsky:1980tq} axion adds a new complex scalar field $\sigma$ and a new heavy quark $Q$ into the standard model. The $U(1)_{PQ}$ transformation is:
\begin{eqnarray}
U(1)_{PQ}:&a&\to a+f_a\alpha\\
&\sigma&\to exp(iq\alpha )\sigma\\
&Q_L&\to exp(iQ\alpha /2)Q_L\\
&Q_R&\to exp(-iQ\alpha/2)Q_R
\end{eqnarray}
where $q$ is the PQ charge of the scalar field $\sigma$. The potential of $\sigma$ is $U(1)_{PQ}$ invariant:
\begin{equation}
V=-\mu_{\sigma}^2\sigma^{*}\sigma+\lambda_{\sigma}(\sigma^*\sigma)^2~~.
\end{equation}
the $U_{PQ}(1)$ symmetry is spontaneously broken by the vav $<\sigma>=v$, therefore:
\begin{equation}
\sigma= (v+\rho)exp(i{a\over v})~~.
\end{equation}
The phase term $a$ acquires a small mass due to non-perturbative effects of QCD instantons thus becomes a pseudo-Goldstone boson. The vav determines the symmetry breaking scale so $f_a=v$, where $f_a$ is the axion decay constant. The heavy quark couples to the Higgs sector as:
\begin{equation}
{\cal L}_{Yu}=-f Q^{\dagger}_L\sigma Q_R-f^*Q^{\dagger}_R\sigma^*Q_R~~.
\end{equation}

The DFSZ \cite{Dine:1981rt} axion on the other hand introduces an additional Higgs doublet instead of a heavy quark. $\sigma$ field couples to the Higgs doublets. The Yukawa terms and the potential terms are:
\begin{eqnarray}
{\cal L}_{Yu}&=&-f_{ij}^{(u)*} q^{\dagger}_{Lj}\phi_2u_{Ri}-f_{ij}^{(d)*} q^{\dagger}_{Lj}\phi_1d_{Ri}+c.c.\\
V&=&(a\phi_1^{\dagger}\phi_1+b\phi_2^{\dagger}\phi_2)\sigma\sigma^*+c(\phi_1^T\phi_2\sigma^2+h.c.)\nonumber\\
&+&d|\phi_1^T\phi_2|^2+e|\phi^T\phi_2|^2+\lambda_1(\phi^{\dagger}_1\phi_1-v_1^2/2)^2\nonumber\\
&+&\lambda_2(\phi^{\dagger}_2\phi_2-v_2^2/2)^2+\lambda(\sigma^*\sigma-v^2/2)^2
\end{eqnarray}
respectively. The $a$, $b$, $c$, $d$ are parameters, $v_{1,2}$ are the vavs of the two Higgs doublets respectively and $i$, $j$ run the family indices. The Peccei-Quinn transformations is:
\begin{eqnarray}
U(1)_{PQ}:~~&\phi_1&\to exp(-i\beta Q)\phi_1\\
&\phi_2&\to exp(-i\gamma Q)\phi_2\\
&u_R&\to exp(i\gamma Q)u_R\\
&d_R&\to exp(i\beta Q)d_R\\
&\sigma&\to exp(iq\alpha )\sigma~,
\end{eqnarray}
where
\begin{eqnarray}
2\alpha=\beta+\gamma~,
\end{eqnarray}
\begin{eqnarray}
\beta={2x\over x+x^{-1}}\alpha~,\\
\gamma={2x^{-1}\over x+x^{-1}}\alpha~,
\end{eqnarray}
and $x=v_2/v_1$.

\section{Axion Properties}
The axion is a pseudo-Goldstone boson which acquires a small mass from QCD instanton effects. The QCD instanton effects after the QCD phase transition can be seen as a mixing between the axion and the pion with a suppression factor which is the symmetry breaking scale factor $f_a$:
\begin{equation}
m_a\sim{m_{\pi^0}f_\pi\over f_a}
\end{equation}
where $f_{\pi}$ is the pion decay constant. The couplings of the axion with the standard model particles can be catalogued into two types: one is due to the ABJ term and the other one is due to mixing. The couplings are suppressed by the symmetry breaking scale $f_a$ as well. For example, the axion couples to the photon and to the standard model fermions as following:
\begin{equation}
{\cal L}_{a\gamma\gamma}\sim {\alpha\over \pi f_a}aF\tilde F~,
\end{equation}
\begin{equation}
{\cal L}_{aq_i}\sim i{m_i\over f_a}a\bar q_i \gamma_5 q_i~,
\end{equation}
where $\alpha$ is the fine structure constant and $m_i$ is the mass of the fermions respectively.

\section{Axion in Cosmology and Astrophysics}
The axion is an attractive candidate for the cold dark matter \cite{Preskill:1982cy,Abbott:1982af,Dine:1982ah,Ipser:1983mw,Stecker:1983,Berezhiani:1990sy,Sikivie:2006ni,Khlopov:1999tm,Valentino:1} as its couplings to the standard model particles are suppressed by the symmetry breaking factor $f_a$ which makes the couplings very weak. In addition, the lifetime of the axion is of order:
\begin{equation}\label{eq:axionLifetime}
T\simeq 10^{50} s\left( \frac{f_a}{10^{12}\,{\rm GeV}} \right)^5~,
\end{equation}
therefore cosmic axions are effectively stable. Finally cosmic axions created from the re-alignment mechanism are very cold and can have the required energy density to be a substantial fraction of dark matter. As the axion field $a$ has an initial value $\theta_0f_a$ where the mis-alignment angle $\theta_0$ is a real number order of one. One can estimate the energy density of the cosmic axions today:
\begin{equation}\label{eq:cdmDensity}
n_a \approx {f^2_a\over 2t_1}({R_1\over R_0})^3,
\end{equation}
where $t_1$ is defined as $m_a(t_1)t_1=1$. The abundance of cosmic axions is therefore:
\begin{equation}
\Omega_a\sim (f_a/10^{12}{\rm GeV})^{7/6}
\end{equation}
which is a substantial fraction of the critical density if $f_a\sim 10^{12}{\rm GeV}$.

The evolutions of stars constrain the axion models as well \cite{Dicus:1978fp,Kuster:2008zz,Raffelt:1987yu,raffelt0607,Dearborn:1985gp,Raffelt:2006,Vysotsky:1978dc,Dicus:1979ch,bahcall,whitedwarf,supernova}. For example, the axions created from the Primakoff effect in stars are an important energy leaking source which changes the evolution of the stars. As the axions are created more abundantly from the horizontal branch stars than from the red giants, the resulted energy leaking decreases the lifetime of the horizontal branch stars more than it does to the red giants. So by comparing the abundance of the red giants and the horizontal branch stars, the axion-photon coupling is constrained:
\begin{equation}
g_{a\gamma\gamma} < 10^{-10}\,\mbox{GeV}^{-1}~~.
\end{equation}

\section{Conclusions}
The introduction of the QCD axion solves the strong CP problem. With a proper PQ symmetry breaking scale $f_a\sim 10^{12}$GeV, the cosmic axions can be a substantial fraction of dark matter.

\section*{Acknowledgments}
This work is supported partially by the Natural Science Foundation of China under Grant Number 11305066 and the Fundamental Research Funds for the Central Universities, HUST: No. 2015TS017. Q. Yang would like to thank Haoran Di for his helpful discussions.

\end{document}